\documentclass[aps,prl,twocolumn, showpacs,reprint,longbibliography,superscriptaddress,10pt]{revtex4-2}
\usepackage[USenglish]{babel}
\usepackage{mathtools}
\usepackage{amssymb}   % for math
\usepackage{amsmath}
\usepackage{placeins}
\usepackage[T1]{fontenc} %to include characters with accents and 
\usepackage[latin1]{inputenc}
\usepackage{lmodern} %to avoid bad quality pdf fonts
\usepackage{color}
\newcommand{\ket}[1]{\left|#1\right\rangle}

\usepackage{float}
\usepackage{graphicx,xcolor}
\usepackage{amsmath, bbm}
\usepackage{hyperref}
\usepackage{epstopdf}
\usepackage{bbm}
\usepackage{soul}
\newcommand{\mean}[1]{\langle #1\rangle}

\renewcommand{\imath}[0]{\mathrm{i}}

 \usepackage[normalem]{ulem}

\begin{document}
\title{Single-molecule strong optomechanical regimes in SERS via hybrid plasmonic cavities}
\author{Miguel \'A. Mart\'inez-Garc\'ia}
\affiliation{Departamento de F\'{i}s\'{i}ca Te\'{o}rica de la Materia Condensada, Universidad Aut\'{o}noma de Madrid, E28049 Madrid, Spain}
\affiliation{Condensed Matter Physics
	Center (IFIMAC), Universidad Aut\'{o}noma de Madrid, E28049 Madrid, Spain}
\author{Johannes Feist}
\author{Diego Mart\'in-Cano}
\email{diego.martin.cano@uam.es}
\affiliation{Departamento de F\'{i}s\'{i}ca Te\'{o}rica de la Materia Condensada, Universidad Aut\'{o}noma de Madrid, E28049 Madrid, Spain}
\affiliation{Condensed Matter Physics
	Center (IFIMAC), Universidad Aut\'{o}noma de Madrid, E28049 Madrid, Spain}
\affiliation{Instituto Nicol\'{a}s Cabrera, Universidad Aut\'{o}noma de Madrid, E-28049 Madrid, Spain}
\date{\today}

\begin{abstract}
We identify enhanced strong coherent Surface-Enhanced Raman Scattering (SERS) interactions facilitated by hybridized metallo-dielectric cavities with a resonant fluorescent molecule. By illuminating a detuned electronic transition near saturation via a narrowband hybrid plasmonic mode, we observe enhanced splittings in the Stokes and anti-Stokes spectra, revealing strong optomechanical coupling and nonlinear vibrational modifications at intensities far below irreversible SERS damage thresholds. We compute the second-order photon correlations that allows identifying the nonclassical character associated to these strong optomechanical interactions and SERS configurations that enables them to optically characterize the anharmonic character of the vibrational modes.
\end{abstract}

\maketitle

A fundamental goal of quantum optics is accessing regimes of strong coupling between light and matter, where the coherent energy exchange prevails over dissipation in the environment, giving rise to rich dynamics and emergent quantum phenomena. 
Some noteworthy examples correspond to the experimental observations of single-emitter strong coupling \cite{Chikkaraddy2016,Pscherer2021}, altered chemical reactivity and modified material properties \cite{Hutchison2012,Feist2018}, as well as cavity and polariton-mediated optomechanical interactions \cite{Verhagen2012, Fainstein2013, Enzian2019}. A field where single-molecule strong coupling seems far from reach is in Raman spectroscopy, where the resulting small dipole moments of molecules compared to the large vibron and cavity decoherences result in very weak interactions. New perspectives arrived from molecular optomechanics \cite{Roelli2016,Schmidt2016,Roelli2024} and its description of Surface Enhanced Raman scattering (SERS) \cite{Itoh2023, Langer2020}, where both the subwavelength plasmonic volumes and near resonant molecular transitions \cite{Neuman2019,Hughes2021} can be potentially exploited  to reach single-molecule strong optomechanical couplings through the increase of the laser driving \cite{Aspelmeyer2014,Koner2023}. However, the available laser intensities that prevent the damage of the samples \cite{Lombardi2018} impose important limitations that make the strong optomechanical coupling regime challenging for SERS. 

\begin{figure}[H]
        \centering
        \includegraphics[width=1.0\linewidth]{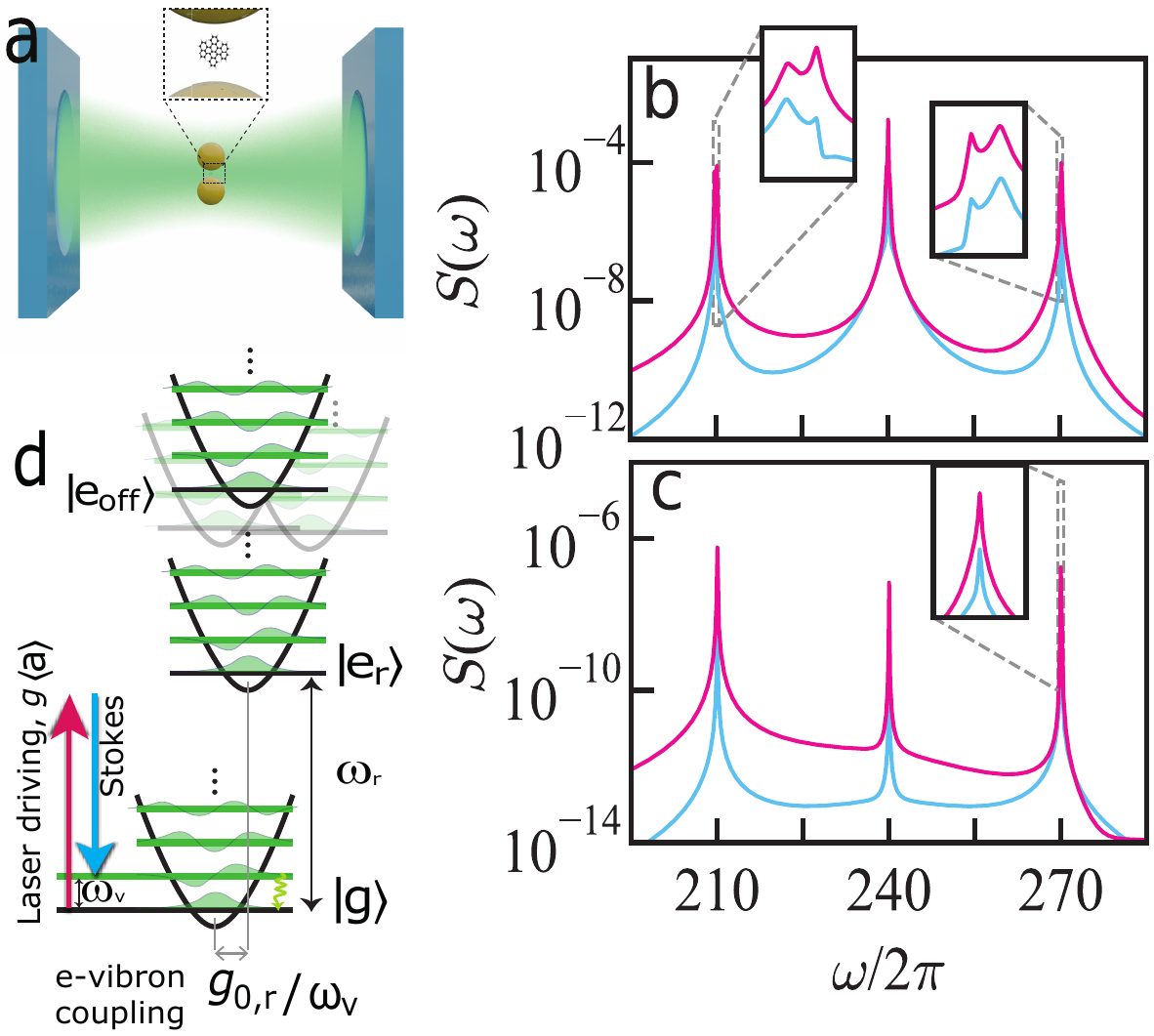}
    \caption{(a) Sketch of a hybrid cavity consisting of a Fabry-Perot resonator coupled to metallic dimer interacting with a molecule in resonance with the anti-Stokes sideband. (b) Cavity spectra of the hybrid system with $g_\mathrm{om}/2\pi=0.1$ THz and $\kappa/2\pi=3.3$ THz ($g_\mathrm{om}/2\pi=0$ for blue lines). (c) Analogous cavity spectra for a metallic dimer without dielectric resonator, i.e. $\kappa/2\pi=33$ THz. The rest of the parameters are $\Delta_c/2\pi=0$, $\omega_v/2\pi=30$ THz, $\Omega/2\pi=1$ THz, $\Delta_r/2\pi=30$ THz, $g_{0,r}/2\pi=3$ THz, $g_r/2\pi=2.5$ THz, $\kappa_v/2\pi=10$ GHz and $\gamma/2\pi=50$ MHz.(d) Sketch of the molecular energy-level structure
for a displaced vibrational potential model with resonant and several off-resonant electronic
levels (black lines)  driven by a resonant laser with the cavity, providing  Stokes and anti-Stokes Raman. The green lines represent the vibrational levels. }
    \label{fig1}
        \end{figure}

In this work we propose a scheme to reach strong electron-vibron coupling in SERS via plasmonic nanocavities hybridized with dielectric resonators \cite{Gurlek2017, Dezfouli2019,Shlesinger2023} as illustrated in Fig.~\ref{fig1}(a). The longer coherence enabled by a hybrid cavity enhances the optomechanical coupling strength compared to a standard plasmonic nanoparticle, allowing it to surpass and modify both vibrational and electronic decoherences. We demonstrate that the frequency splitting of the peak due to strong coupling is associated with nonclassical photon statistics, which displays a clear bunching correlation between Stokes and anti-Stokes peaks. Finally, we show a SERS configuration that enables exploiting the strong coupling correlations in the Stokes signals to characterize the nonclassical features of molecular anharmonic vibrations through photon statistics.

A first-principles SERS model for a single vibration can be derived from a  Hamiltonian that accounts for the electron-vibron interactions between resonant and off-resonant levels \cite{Tommasini2009,Martinez-Garcia2024} (see sketch in Fig.\ref{fig1}(d)). For a single nearly resonant transition and vibration with annihilation operators $\hat{\sigma}_r$ and $\hat{v}$, the resulting Hamiltonian can be simplified in the rotating frame of the laser at frequency $\omega_L$ as \cite{Martinez-Garcia2024}
\begin{align}
	\label{eqn:AdiabaticEliminationHamiltonian}
	H&= \hbar \Delta_c \hat{a} ^\dagger\hat{a}+ \hbar\omega_v \hat{v}^\dagger \hat{v}+\hbar\Omega(\hat{a}^\dagger + \hat{a})+\hbar g_{\mathrm{om}} \hat{a}^\dagger \hat{a} (\hat{v}^\dagger + \hat{v})\\ \nonumber
	&+\hbar \Delta_r \hat{\sigma}_r ^\dagger\hat{\sigma}_r+ \hbar g_{0,r} \hat{\sigma}_r^\dagger \hat{\sigma}_r (\hat{v}^\dagger + \hat{v})+ \hbar g_r (\hat{\sigma}_r^\dagger  \hat{a}+ \hat{\sigma}_r \hat{a}^\dagger ),
\end{align}
where $g_{0,r}$ denotes the electron-vibrational coupling and $g_{\mathrm{om}}$ is the optomechanical Raman interaction strength arising from off-resonant transitions dispersively coupled to a hybrid plasmonic mode characterized by the annihilation operator $\hat{a}$. The cavity is driven by a laser field with Rabi frequency $\Omega$ and frequency $\omega_L$. The laser detunings  with respect to the optical transitions and the plasmon resonance are given by $\Delta_r = \omega_r - \omega_L$ and $\Delta_c = \omega_c - \omega_L$, respectively. We consider that the optical electronic transition is close to resonance with a single hybrid cavity mode with coupling strength $g_r$. 
Furthermore, we adopt an open-system approach~\cite{Gardiner2004}, in which each cavity, vibrational, and electronic degree of freedom couples to distinct environmental baths. For typical molecular parameters, the coupling strengths are smaller than the bare transition frequencies, which justifies a Markovian description of their decay into the baths. This scenario is modeled using a standard master equation~\cite{Gardiner2004} $
	\dot{\hat\rho} = -i[\hat{H}, \hat\rho] + \gamma \mathcal{L}_{\hat{\sigma}_r}[\hat{\rho}] + \kappa \mathcal{L}_{\hat{a}}[\hat{\rho}] + \kappa_v \mathcal{L}_{\hat{v}}[\hat{\rho}]$,
where the Lindblad superoperator is defined as
$
	\mathcal{L}_{\hat{O}}[\hat{\rho}] = \hat{O} \hat{\rho} \hat{O}^\dagger - \frac{1}{2}\{\hat{O}^\dagger \hat{O},\hat{\rho}\}.
$
This formulation captures the dissipative processes due to spontaneous emission with rates $\gamma$, vibrational relaxation with rate $\kappa_v$, and cavity losses characterized by $\kappa$.

Figure~\ref{fig1}(b) displays the steady-state cavity emission spectra ($S(\omega) = \frac{1}{\pi} \text{Re} \int_{0}^{\infty} \lim_{t \to \infty} \langle \hat{a}^\dagger(t ) \hat{a}(t+ \tau) \rangle e^{i \omega \tau} d\tau$) for a narrowband hybrid cavity mode in resonance with the laser, and the optical molecular transition at the anti-Stokes frequency. These narrowband hybrid modes arise from the interference between a photonic resonator and a broadband  plasmonic mode of a nanoparticle \cite{Doeleman2016,Gurlek2017,Dezfouli2017a} resulting from the interaction between the nanoparticle polarizability and the resonator, and enable access to the resolved-sideband regime of optomechanics~\cite{Shlesinger2023} ($\kappa_\mathrm{hyb}<\omega_v\lesssim \kappa_\mathrm{pl}$). More importantly, the increased interaction via the resonator enables to enhance the Purcell factor $F$ relative to that of the bare nanoparticle, thereby enhancing the coupling-to-loss ratio $g_r/\kappa$ crucial for resonant SERS  ($g_\mathrm{hyb}/\kappa_\mathrm{hyb}=\sqrt{\frac{F_\mathrm{hyb}\kappa_\mathrm{pl}}{F_\mathrm{pl}\kappa_\mathrm{hyb}}}g_\mathrm{pl}/\kappa_\mathrm{pl}\sim 10g_\mathrm{pl}/\kappa_\mathrm{pl}$, consistent with the values reported in the literature \cite{Doeleman2016,Gurlek2017}, i.e. $F_\mathrm{hyb}/F_\mathrm{pl}\sim 3-7$, $\kappa_\mathrm{pl}/\kappa_\mathrm{hyb}\sim100-10$). Therefore for these simulations we have considered typical molecular and nanocavity parameters similar to those in previous works~\cite{Martinez-Garcia2024,Roelli2016}, but as opposed to the results there presented, by considering the narrowband hybrid linewidth \cite{Gurlek2017} we observe a clear splitting in the Stokes and anti-Stokes transitions that are absent when the cavity is replaced by a broader plasmonic resonance (Figure~\ref{fig1}(c)).  Similar phenomena have been reported in the presence of strong driving in resonance with both the emitter and a broad plasmonic cavity~\cite{Neuman2019}, where the resonant Mollow triplet transitions match the Raman sidebands and become dressed via strong electron-vibron coupling. As an important difference, in this work we are considering a detuned emitter-hybrid cavity configuration that gives access to the strong optomechanical electron-vibron interactions at driving powers that are well below the sample damaging limits within SERS experiments \cite{Lombardi2018,Roelli2016,Zrimsek2016} ($\Omega/2\pi=\sqrt{\kappa}\sqrt{P_\mathrm{in}/\hbar\omega_L}/2\pi\sim 1-10 $ THz, corresponding approximately to a CW pump power $P_\mathrm{in}$ range from $1$ to $10~\mu$W).

\begin{figure*}
	\centering
	\includegraphics[width=1.0\textwidth, height=7cm]{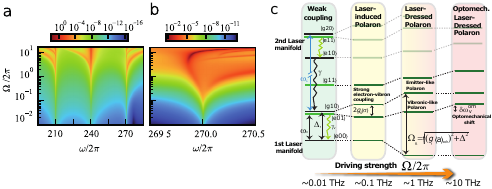}
	\caption{(a) Cavity emission spectra as a function of the driving strength  and (b) a frequency zoom at the anti-Stokes region. The parameters coincide with Fig.~1.   (c) Optomechanical laser-dressed polaron energy level scheme in nearly resonant SERS  as a function of the driving strength $\Omega$ and with off-resonant cavity optomechanical interactions. The left panel illustrates the laser-detuned energy structure in the weak electron-vibron coupling region, which is shown in the bare basis of the electronic, laser and vibrational excitation-number states, $\ket{n_e,n_L,n_v}$. The right panels display the three distinctive polaron interaction regions as a function of the driving field. }
\label{fig2}
\end{figure*}

To further understand the origin of this strong coupling phenomenon, in Fig.~\ref{fig2}(a)  we represent the spectral map of the Stokes and anti-Stokes sidebands as a function of the laser driving strength, which display four interaction regions. For weak drivings, the Raman lines are uniquely identified with the vibrational frequency, but at a critical driving the line becomes increasingly split in two due to the coherent driving of the hybrid mode (see zoom of the anti-Stokes lines in panel (b)). These new eigenenergies arise as a result of the strong optomechanical coupling between the electron and vibron, which creates two laser-induced polarons (see the two left panels in Fig.~\ref{fig2}(c)). In this picture, the accumulated photons in the hybrid cavity enhances the resonant coherent interaction of the Raman sideband with the detuned emitter zero-phonon line (ZPL), thus facilitating the entrance of the strong coupling regime. This configuration avoids the higher laser intensities involved in resonant illuminations of the ZPL, where the Mollow triplet sidebands must match the Raman transitions ($\Omega \approx\kappa\omega_v/2g_r\gtrsim \omega_v$)~\cite{Neuman2019}.  The strong coupling arises when the cavity-dressed resonant optomechanical coupling (2$g_{0,r}\lvert \mean{\sigma }\lvert\approx 4g_{0,r}g_r\Omega/\kappa\Delta_r $) surpasses the Purcell-enhanced electronic and vibron decoherence rates \cite{Ramos2013} ($2g_r^2 \kappa/(\kappa^2+\Delta_c^2)$ and $\gamma_v/2$, respectively), as inferred from a simplified linearized optomechanical model where the cavity is adiabatically eliminated.

Interestingly, at stronger drivings, we further notice that the laser-induced polaron lines in both the Stokes and anti-Stokes sidebands become nonlinearly shifted and broadened with respect to the conventional vibrational frequency and decay rate (see panels (a) and (b)). Contrary to most scenarios in nonresonant SERS where the vibrational lines remain unaltered~\cite{Roelli2024}, we emphasize that the enhanced optomechanical cooperativity of the resonant transition and the hybrid cavity enables significant vibrational modifications of the polarons, showing the promise to observe strong optomechanical nonlinear effects at the single-molecule level, which differ from perturbative nonlinear optical processes \cite{Chen2021,Xomalis2021}. 
As opposed to the harmonic character of standard optomechanical systems~\cite{Aspelmeyer2014}, the saturable nature of the emitter transition enables different strong nonlinear vibrational modifications that can be explored in nearly resonant SERS. The emitter component of the laser-induced polaron eigenstates, analogous to laser-dressed polar mechanisms \cite{Groiseau2024}, thus becomes further dressed with the laser with a Rabi frequency $\Omega_R =\sqrt{\Delta_r^2+(2g_r\Omega/\kappa)^2}$, which is provided by the resonantly-fed coherent cavity amplitude ($|\mean{\hat{a}_\mathrm{coh}}|=2g_r\Omega/\kappa$). This results in intra and interband Mollow triplet transitions \cite{Groiseau2024} assisted by resonant interactions with the vibrations, which are  modified by the off-resonant optomechanical Raman coupling at the strongest drivings, where it creates relevant vibrational shifts (see the two right panels in Fig.~\ref{fig2}(c)). These complex nonlinear regions can be fairly reproduced by a simpler Hamiltonian, arising from the interplay between a linearized cavity optomechanical interaction Hamiltonian and a cavity-dressed polaron one. The latter one is obtained by first adiabatically eliminating the cavity and then applying a laser-dressed transformation \cite{Groiseau2024} that accounts for the emitter's saturation, followed by a zeroth-order polaron transformation \cite{Fur2026}. 

The resulting spectra therefore show nonmonotonic vibrational modifications with the driving strength that differ with respect to linearized optomechanical cooling and spring effects formerly reported~\cite{Aspelmeyer2014,Neuman2019}, which result from a complex combination of resonant and off-resonant optomechanical couplings and the AC Stark effect of the ZPL, in which the resonant terms dominate at lower drivings and the off resonant Hamiltonian dominate at higher ones. At strong drivings, we also observe additional transitions in which further Raman sidebands split from the vibrational lines. These nonlinear features arise from higher-order vibrational levels that become dressed with the exciton due to the longer coherence induced by the narrow hybrid cavity mode, without the need of reaching ultrastrong electron-vibron couplings \cite{Neuman2019}, and are further enhanced with ZPLs at Stokes-detuned configurations (not shown).

\begin{figure}
        \centering
        \includegraphics[width=1.0\linewidth]{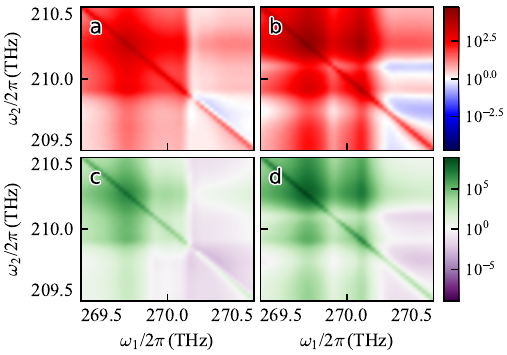}
    \caption{Maps of the normalized frequency-filtered second-order coherence $g_\Gamma^{(2)}(\omega_1,\omega_2)$ at two frequencies for the purely resonant model (a) and (b) with off-resonant levels and their corresponding Cauchy-Schwarz inequality \cite{SanchezMunoz2014} $( g_\Gamma^{(2)}(\omega_1,\omega_2) )^2 /( g_\Gamma^{(2)}(\omega_1,\omega_1)g_\Gamma^{(2)}(\omega_2,\omega_2) )$ maps , (c)-(d), respectively . The rest of the parameters coincide with Fig.~1. }
    \label{fig3}
        \end{figure}

We continue evaluating the photon statistics in order to quantify the quantum optomechanical correlations and identify distinctive features arising from the strong electron-vibron coupling. By means of the sensor method \cite{SanchezMunoz2014}, we compute in Fig.~\ref{fig3} the second-order correlation function filtered at two different frequencies near the Stokes (y-axis) and anti-Stokes transitions (x-axis). To assess the role of off-resonant electronic states in strong Raman scatterers \cite{Martinez-Garcia2024}, we compute the spectra for a single resonant Raman model (a) and for a cooperative electronic model with off-resonant levels (b). Beside the strongly bunched diagonal line common in optomechanical systems \cite{Schmidt2021}, we identify two more spots in both the diagonal and its perpendicular direction, forming a four spots square pattern. Similarly to our previous results in weakly optomechanically coupled systems \cite{Martinez-Garcia2024}, the cooperative model (b) shows an enhancement of the photon correlations displaying more prominent features with respect to the single resonant one, but as a crucial difference, the correlated spots in both models  only arise in the presence of strong optomechanical coupling. These superbunching spots are a result of the two-photon transitions providing a Stokes and anti-Stokes photon, which are assisted by the two laser-induced polaron resonances arising from the strong optomechanical coupling. This enhanced photon pair production is a result of  nonclassical quantum correlations as reflected in the violations of the Cauchy-Schwarz inequality \cite{SanchezMunoz2014} that display values orders of magnitude above one, see panels (c) and (d) respectively.

The enhanced production of Stokes and anti-Stokes photons via the resolved-sideband character of the hybrid cavity and the ZPL can enable novel types of SERS correlation spectroscopy. This photonic spectroscopy can allow us to gain information about the statistics of the molecular vibrations, which otherwise is challenging to access. Recent works have shown that the anharmonic character of the molecular vibrations can affect both, the correlations of vibrations \cite{Schmidt2024} and photons in optomechanical up-conversion \cite{Kalarde2025}. In the following we explore the influence of vibrational anharmonicity in SERS correlation spectroscopy by computing the second-order Stokes-Stokes correlation spectra in the presence of a nonlinear vibrational potential term in the Hamiltonian, $\frac{\eta \omega_v}{2} \hat{v}^\dagger  \hat{v}^\dagger  \hat{v} \hat{v}$, where $\eta$  parametrizes the degree of anharmonicity of the vibrational mode. 

\begin{figure}
	\centering
	\includegraphics[width=1.0\linewidth]{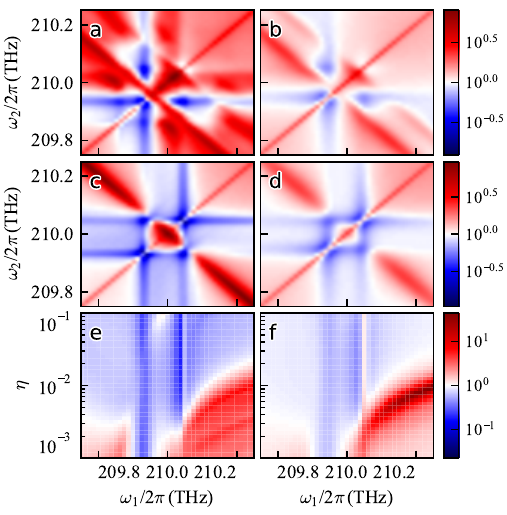}
	\caption{ Maps of the normalized frequency-filtered second-order coherence $g_\Gamma^{(2)}(\omega_1,\omega_2)$ at two frequencies for the combined model and a cavity linewidth for an harmonic vibration ($\eta=0$) for a hybrid $\kappa/2\pi=3.3$ THz (a) and plasmonic nanocavity $\kappa/2\pi=33$ THz (b). (c)-(d) analogous figuresfor an anharmonic vibration ($\eta=0.05$) . (e)-(f) the second sensor frequency is fixed at $\omega_2/2\pi=210.05$ THz as a function the anharmonicity parameter $\eta$. The respective parameters are $\Delta_c^\prime=-\omega_v$ and $\Omega/2\pi=7$ THz. The rest of the parameters coincide with Fig.~1. }
	\label{fig4}
\end{figure}
Figure~\ref{fig4}(c) shows the two-photon correlation spectra for an anharmonic vibrational mode in the frequency range near the Stokes transitions, which display the largest Raman spectral signals in SERS. To increase the optical production of vibrational excitations, together with strong electron-vibron interaction, we set the cavity resonance at the Stokes transition and the ZPL at the anti-Stokes one. The strongly antibunched square pattern (panel (c)) that results from the dressed polaron resonances displays an upper right vertex in the antidiagonal with nonclassical statistics that is absent for an harmonic vibration, which displays a superbunched peak instead (compare panels (a) and (c)). The resulting antibunched two-photon correlation map is analogous to that of two interacting two-level systems \cite{Gonzalez-Tudela2013}, which emphasizes the nonclassical single-particle character of the vibration  provided by the anharmonic potential. Similar behaviours are also achieved by a  broadband plasmonic resonance (see panels (b) and (d), respectively),  since the detuned cavity illumination configuration  makes both cavity coherences similar due to the relevant weight of the detuning ($\propto g_r/ (\Delta'^2_c+\kappa^2)$). Nevertheless, the narrower hybrid resonance configuration still displays more pronounced nonclassical features compared to the broadband plasmonic cavity due to its enhanced coherence, which together with its stronger interacting capabilities highlights the potential of hybrid cavities for exploring strong optomechanical anharmonicities in resonant SERS. Despite the generality of the detuned nanocavity-driving configurations to explore anharmonicity via Stokes photon statistics in resonant SERS, notice that they require a stronger laser intensity to compensate the weaker Raman interactions compared to the resonant hybrid mode configuration (i.e. results in Fig.~\ref{fig3}), which is still below the maximum drivings accessible in current experiments \cite{Roelli2016,Lombardi2018}. 

To explore the influence of the anharmonic character of the molecular vibration on the degree of photon antibunching in SERS, in Fig.~\ref{fig4}(e) and (f) we plot the second-order photon correlation as a function of the anharmonicity and the first sensor frequency, setting the second sensor at the strongly antibunched frequency of the vibron-like polaron in panels (c) and (d) ($\omega_2/2\pi=210.05$ THz). We can observe a clear onset of  antibunching at stronger anharmonicities for both cavities. Only at the diagonal frequency region ($\omega_1/2\pi\sim 209.95$ THz), the antibunching is independent of the anharmonicity degree of the molecular potential, which results from the impossibility to detect two Stokes photons resulting from both polaron resonances.  Nevertheless, the strong antibunching shown at the antidiagonal region ($\omega_1/2\pi\sim 210.05$ THz) shows the potential to explore the anharmonic nature of the molecular vibrations via photon correlations for both hybrid \cite{Shlesinger2023} and standard SERS nanocavities \cite{Zrimsek2016}. We notice that we have chosen sufficiently narrowband filters to resolve the spectral features of strong coupling, but stronger degrees of antibunching with broader filters can be achieved (not shown), similar to the case of optomechanical up-conversion~\cite{Kalarde2025}.

To conclude, we have shown that hybrid nanocavities can facilitate reaching the single-molecule strong electron-vibron coupling regimes, opening a promising route to explore their features in resonant SERS under feasible experimental conditions. In particular, they enable enhanced spectral splittings in resonant Raman responses at microwatt incident powers, which are several orders of magnitude below the milliwatt thresholds preventing damage in standard experiments. The cooperative enhancement between resonant and off-resonant contributions for strong Raman molecular scatterers manifests directly in more intense signals of the observed spectral splittings, which can turn into a key ingredient for the observation of strong optomechanical coupling in SERS. Moreover, we have shown  that  the resonant and off-resonant optomechanical interaction leads to modifications of the laser-induced polaronic vibrational frequencies and decay rates at high but reachable intensities, which provides a roadmap to explore purely optomechanical dynamical effects in resonant SERS beyond perturbative nonlinearities shown in optical conversion \cite{Chen2021,Xomalis2021}. These features are reflected in the generation of highly correlated pairs of Stokes-anti-Stokes photons at the laser-induced polaron frequencies, which can show orders of magnitude violations of the Cauchy-Schwartz inequality. 
We have also proposed a method to exploit the strong optomechanical coupling to characterize the degree of vibrational nonlinear potentials through the measurement of photon antibunching in the scattered Raman fields. This approach provides an optical signature of nonlinear vibrational potentials and lays the groundwork for the optical exploration of higher order molecular nonlinearities. Altogether, these advances outline a roadmap toward quantum optomechanics of individual molecules, where vibrational degrees of freedom can be interrogated and controlled photon by photon. In this regard, the higher optical coherences provided by hybrid nanostructures compared to broadband plasmonic nanostructures suggests the research of novel architectures that may open the search of quantum-coherent optomechanical interactions at the single-photon level at experimentally accessible conditions. In such regimes, one may envision tracking vibrational quantum trajectories via photon statistics, effectively using photons as a probe to mimic and reconstruct vibrational dynamics with optical means \cite{Ludwig2012,Aspelmeyer2014}. This capability would bridge quantum optics and molecular SERS spectroscopy, opening new opportunities to investigate vibrational coherence and dissipation at a fundamental level.

\begin{acknowledgments}
 We acknowledge financial support by the Spanish Ministerio de Ciencia y Universidades-Agencia Estatal de Investigaci\'on through Grants No. PID2021-126964OB-I00, No. TED2021-130552B-C21, No. PID2024-156077OB-I00, No. PID2024-161142NB-I00, and No. CEX2023-001316-M through the Mar\'ia de Maeztu program for Units of Excellence in RD, and from the European Union's Horizon Europe research and innovation programme under Grant Agreements No. 101070700 (MIRAQLS) and No. 101098813 (SCOLED). The authors acknowledge the support by the project NPhoQuss funded by Programa Fundamentos 2024 from the BBVA Foundation through the grant EIC24-1-17304.
\end{acknowledgments}

\end{document}